\documentclass[useAMS,referee,usenatbib]{biom}
\usepackage{natbib}
\usepackage{url}
\usepackage{bm}
\usepackage[flushleft]{threeparttable}
\usepackage{booktabs}
\usepackage{rotating}
\usepackage{adjustbox}
\usepackage{xcolor}
\usepackage{changepage}
\usepackage{verbatim}
\usepackage{amsmath}
\usepackage{amsfonts}
\usepackage{graphicx}
\usepackage{tikz}
\usepackage{multirow}
\usepackage{subcaption}
\usepackage{bigints}
\usepackage{paralist}
\usepackage{rotating}

\usepackage{hyperref}
\hypersetup{colorlinks=true, linkcolor=blue, citecolor=blue, filecolor=blue}

\usepackage{enumerate}
\usepackage[medium,compact]{titlesec}
\allowdisplaybreaks

\def\bSig\mathbf{\Sigma}

\newtheorem{proposition}{Proposition}

\newcommand{\bcx}{{\bm X}}

\newcommand{\bcy}{{\bm Y}}
\newcommand{\bcv}{{\bm V}}

\newcommand{\bco}{{\bm O}}

\newcommand{\bcm}{{\bm M}}

\newcommand{\bcc}{{\bm C}}

\newcommand{\bmm}{{\bm m}}

\newcommand{\E}{\mathbb{E}}
\newcommand{\Prob}{\mathbb{P}}

\newcommand{\differential}{\text{d}}

\allowdisplaybreaks

\usetikzlibrary{shapes,decorations,arrows,calc,arrows.meta,fit,positioning}
\tikzset{
    -Latex,auto,node distance =1 cm and 1 cm,semithick,
    state/.style ={ellipse, draw, minimum width = 0.7 cm},
    point/.style = {circle, draw, inner sep=0.04cm,fill,node contents={}},
    bidirected/.style={Latex-Latex,dashed},
    el/.style = {inner sep=2pt, align=left, sloped}
}

\title[Causal mediation in cluster-randomized trials]{Semiparametric causal mediation analysis of cluster-randomized trials for indirect and spillover effects}

\author{Chao Cheng$^{1,*}$\email{chaoc@wustl.edu} and Fan Li$^{2,3,**}$\email{fan.f.li@yale.edu} \\
$^{1}$Department of Statistics and Data Science, Washington University in St. Louis, St. Louis, MO, USA \\
$^{2}$Department of Biostatistics, Yale School of Public Health, New Haven, CT, USA \\
$^{3}$Center for Methods in Implementation and Prevention Science, Yale School of Public Health,\\
New Haven, CT, USA 
}

\begin{document}





\pagerange{\pageref{firstpage}--\pageref{lastpage}} 
\volume{64}
\pubyear{2008}
\artmonth{December}


\doi{10.1111/j.1541-0420.2005.00454.x}


\label{firstpage}


\begin{abstract}
In cluster-randomized trials (CRTs), there is emerging interest in exploring the causal mechanism in which a cluster-level treatment affects the outcome through an intermediate outcome. The majority of existing causal mediation methods are applicable to independent data and only a few exceptions have considered assessing causal mediation in CRTs, all of which heavily depend on parametric assumptions. In this article, we develop a formal semiparametric efficiency theory to motivate new doubly-robust methods for addressing different mediation effect estimands---the natural indirect effect, individual mediation effect, and spillover mediation effect (the extent to which one's outcome is influenced by others' mediators). We derive the efficient influence function for each estimand, and carefully parameterize each efficient influence function to motivate practical estimators. We consider both parametric working models and data-adaptive machine learners to estimate the nuisance functions, and obtain the semiparametric efficient estimators in the latter case. We conduct simulation studies to demonstrate the finite-sample performance of our new estimators and illustrate our proposed methods by reanalyzing a real-world CRT. 
\end{abstract}

%



\maketitle


%


\section{Introduction}
Cluster-randomized trials (CRTs) are common in health and social sciences to study the population-level treatment effect. While the total treatment effect has been a cornerstone in the analysis of CRTs, there is an emerging interest for conducting mediation analysis in understanding the mechanisms by which a cluster-level treatment affects the outcome. 
By decomposing the total effect into a natural indirect effect (NIE) through the mediator and a natural direct effect (NDE) bypassing the mediator, mediation analysis holds the promise to advance the theory underlying process evaluation and optimizing future interventions. 

To assess mediation in CRTs, a few methods have been developed to address within-cluster correlation and interference \citep{hudgens2008toward}. For example, 
\citet{park2015bayesian} provided a set of identification conditions for the NIE, and employed a Bayesian multilevel modeling approach for estimation. 
Their approach implicitly assumes no within-cluster interference, and imposes strong parametric modeling assumptions. Relaxing the no interference assumption, \citet{vanderweele2010direct} and \citet{vanderweele2013mediation} provided a further decomposition of the NIE into a \emph{spillover mediation effect} (SME) and an \emph{individual mediation effect} (IME), for which identification conditions and nonparametric identification formulas are provided. Each identification formula permits the use of multilevel models to derive the mediation effects, and the consistency of the final estimator critically depends on the correct specification of the fitted multilevel models. Furthermore, the existing methods have also assumed away \emph{informative cluster size}, whereby the cluster size may be a surrogate of the within-cluster dynamics that is predictive of the mediator and/or the outcome and hence should be an intrinsic element of the CRT estimands \citep{kahan2023estimands}.

This article formalizes a semiparametric approach to assess mediation in CRTs that addresses the limitation of the previous developments. First, we operate under the general setup of informative cluster size \citep{cruces2025design,davezies2009identification,jiang2023statistical,kahan2023estimands}, which gives rise to two versions of the total effect estimands. The \textit{cluster-average treatment effect} targets the average change on the population of clusters along with their natural cluster members, whereas the \textit{individual-average treatment effect} targets the average change on the population of all individuals across clusters. Based on the two versions of the total effect estimands, the mediation effect estimands can be defined at both the cluster and individual level. 
Second, we account for within-cluster interference and provide causal assumptions to point identify NIE, NDE, as well as SME that quantifies the extent to which one’s outcome is influenced by others' mediators. 
Leveraging the semiparametric efficiency theory \citep{bickel1993efficient}, we derive the efficient influence function (EIF) of each estimand, and characterize the optimal estimator with and without reparameterization. Finally, the proposed estimators are further enhanced by leveraging machine learning tools for nuisance estimation under a cross-fitting scheme \citep{chernozhukov2018double}. 

\section{Preliminaries}\label{sec:preliminaries}
\subsection{Notations and data structure}
We consider a CRT with $K$ clusters. For cluster $i$, we define $N_i$ as the cluster size,  $A_i \in \{0,1\}$ as the cluster-level treatment, and $\bcv_i \in \mathbb{R}^{d_{\bcv}\times 1}$ as a vector of cluster-level baseline covariates. For individual $j$ of cluster $i$, let $\bcx_{ij} \in \mathbb{R}^{d_{\bcx}\times 1}$ be a vector of individual-level baseline covariates, and write $\bcx_i=[\bcx_{i1},\dots,\bcx_{iN_i}]^T\in \mathbb{R}^{N_i\times d_{\bcx}}$.  Let $\bcc_i = \{\bcv_i,\bcx_i\}$ be all baseline covariates in cluster $i$. 
We define $Y_{ij} \in \mathbb{R}$ as the individual-level outcome, $M_{ij} \in \mathbb{R}$ as the individual-level mediator that is measured before the outcome but after treatment. We define $\bcy_i = [Y_{i1},\dots,Y_{iN_i}]^T\in \mathbb{R}^{N_i\times 1}$, $\bcm_i = [M_{i1},\dots,M_{iN_i}]^T\in \mathbb{R}^{N_i\times 1}$, and $\bcm_{i(-j)} \in \mathbb{R}^{(N_i-1)\times 1}$ as a vector of mediators from cluster $i$ excluding individual $j$. To summarize, we observe $\bco_i = \{N_i,\bcc_i,A_i,\bcm_i,\bcy_i\}$, $i=1,\dots,K$, where causal  relationships among these variables are illustrated in Figure \ref{fig:dag}(a).  For conciseness, we sometimes omit the cluster indicator $i$ in the subscript, such that $\bm O$ for $\bm O_i$, $Y_{\cdot j}$ for $Y_{ij}$, $M_{\cdot j}$ for $M_{ij}$, and $\bcm_{\cdot(-j)}$ for $\bcm_{i(-j)}$.

We define $M_{\cdot j}(a)$ as the potential mediator 
under $a\in\{0,1\}$, $\bcm(a)=[M_{\cdot 1}(a),\dots,M_{\cdot N}(a)]^T$ as the vector of potential mediators, and $\bcm_{\cdot (-j)}(a)$ as the vector by excluding the $j$th element in $\bcm(a)$. Define $Y_{\cdot j}(a,\bmm)$ as the potential outcome when treatment is set to $a$ and the mediators of all individuals in that cluster, $\bcm$, are set to value $\bmm$. One can equivalently represent $Y_{\cdot j}(a,\bmm) = Y_{\cdot j}(a,m_{\cdot j},\bmm_{\cdot (-j)})$ with $\bmm=\{m_{\cdot j},\bmm_{\cdot (-j)}\}$; this notation explicitly distinguishes an individual's own mediator from the mediators of the remaining cluster members. We also assume $Y_{\cdot 
 j}(a)=Y_{\cdot j}\left(a,\bcm(a)\right)=Y_{\cdot j}\left(a,M_{\cdot j}(a),\bcm_{\cdot (-j)}(a)\right)$; i.e., the potential outcome under $A=a$ is identical to the potential outcome when $A=a$ and all mediators in that cluster are set to their natural values under treatment $a$. 

In CRTs, cluster size can be associated with both mediator and outcome, leading to informative cluster size. 
Without ruling out informative cluster size, the total treatment effect can be measured by either
the \textit{cluster-average treatment effect} or \textit{individual-average treatment effect} (\citealp{wang2023model}): $\text{TE}_C = g\left(\mu_C(1),\mu_C(0)\right)$ and $\text{TE}_I = g\left(\mu_I(1),\mu_I(0)\right)$, 
where $g(\cdot,\cdot)$ is a function determining the scale of effect measure, and
\begin{equation*}
\mu_C(a) = \E\left\{\frac{\sum_{j=1}^{N}Y_{\cdot j}(a)}{N}\right\}, \quad \mu_I(a) = \frac{\E\left\{\sum_{j=1}^{N}Y_{\cdot j}(a)\right\}}{\E\{N\}},~~~a\in\{0,1\}. 
\end{equation*}
For example, $g(x,y)=x-y$, $g(x,y)=\frac{x}{y}$ and $g(x,y)=\frac{x/(1-x)}{y/(1-y)}$ correspond to causal mean difference, causal risk ratio and causal odds ratio, respectively. $\mu_C(a)$ represents the average potential outcome under $A=a$ for the population of clusters, whereas $\mu_I(a)$ represents the average potential outcome under $A=a$ among the population for all individuals across clusters. Intuitively, $\text{TE}_I$ resembles a natural estimand that one would have targeted under individual randomization, but $\text{TE}_C$ is more specialized to cluster randomization.

\begin{figure}[htbp]
\centering
\begin{minipage}{.30\linewidth}
\subfloat[all causal pathways]{
\begin{tikzpicture}[scale=0.55, transform shape]
    \node[state] (a) at (0,0) {$A_i$};
    \node[state] (m1) at (3,3) {$M_{i1}$};
    \node[state] (m2) at (3,0) {$M_{i2}$};
    \node[state] (m3) at (3,-3) {$M_{i3}$};
    \node[state] (y1) at (6,3) {$Y_{i1}$};
    \node[state] (y2) at (6,0) {$Y_{i2}$};
    \node[state] (y3) at (6,-3) {$Y_{i3}$};
    \draw (a) edge (m1);
    \draw (a) edge (m2);
    \draw (a) edge (m3);
    \draw (a) edge[bend left=65] (y1);
    \draw (a) edge[bend left=-25] (y2);
    \draw (a) edge[bend left=-65] (y3);
    \draw (m1) edge (y1);
    \draw (m2) edge (y2);
    \draw (m3) edge (y3);
    \draw (m1) edge (y2);
    \draw (m1) edge (y3);
    \draw (m2) edge (y1);
    \draw (m2) edge (y3);
    \draw (m3) edge (y1);
    \draw (m3) edge (y2);
    \draw [gray,dashed,-] (m1) edge (m2);
    \draw [gray,dashed,-] (m2) edge (m3);
    \draw [gray,dashed,-] (m1) edge[bend left=-30] (m3);
    \draw [gray,dashed,-] (y1) edge (y2);
    \draw [gray,dashed,-] (y2) edge (y3);
    \draw [gray,dashed,-] (y1) edge[bend left=30] (y3);
\end{tikzpicture}}
\end{minipage}
\begin{minipage}{.30\linewidth}
\subfloat[natural direct effect]{
\begin{tikzpicture}[scale=0.55, transform shape]
    \node[state] (a) at (0,0) {$A_i$};
    \node[state] (m1) at (3,3) {$M_{i1}$};
    \node[state] (m2) at (3,0) {$M_{i2}$};
    \node[state] (m3) at (3,-3) {$M_{i3}$};
    \node[state] (y1) at (6,3) {$Y_{i1}$};
    \node[state] (y2) at (6,0) {$Y_{i2}$};
    \node[state] (y3) at (6,-3) {$Y_{i3}$};
    \draw (a) edge[bend left=65] (y1);
    \draw (a) edge[bend left=-25] (y2);
    \draw (a) edge[bend left=-65] (y3);
    \draw [gray,dashed,-] (m1) edge (m2);
    \draw [gray,dashed,-] (m2) edge (m3);
    \draw [gray,dashed,-] (m1) edge[bend left=-30] (m3);
    \draw [gray,dashed,-] (y1) edge (y2);
    \draw [gray,dashed,-] (y2) edge (y3);
    \draw [gray,dashed,-] (y1) edge[bend left=30] (y3);
\end{tikzpicture}}
\end{minipage}
\begin{minipage}{.30\linewidth}
\subfloat[natural indirect effect]{
\begin{tikzpicture}[scale=0.55, transform shape]
    \node[state] (a) at (0,0) {$A_i$};
    \node[state] (m1) at (3,3) {$M_{i1}$};
    \node[state] (m2) at (3,0) {$M_{i2}$};
    \node[state] (m3) at (3,-3) {$M_{i3}$};
    \node[state] (y1) at (6,3) {$Y_{i1}$};
    \node[state] (y2) at (6,0) {$Y_{i2}$};
    \node[state] (y3) at (6,-3) {$Y_{i3}$};
    \draw (a) edge (m1);
    \draw (a) edge (m2);
    \draw (a) edge (m3);
    \draw [white] (a) edge[bend left=65] (y1);
    \draw [white] (a) edge[bend left=-25] (y2);
    \draw [white] (a) edge[bend left=-65] (y3);
    \draw (m1) edge (y1);
    \draw (m2) edge (y2);
    \draw (m3) edge (y3);
    \draw (m1) edge (y2);
    \draw (m1) edge (y3);
    \draw (m2) edge (y1);
    \draw (m2) edge (y3);
    \draw (m3) edge (y1);
    \draw (m3) edge (y2);
    \draw [gray,dashed,-] (m1) edge (m2);
    \draw [gray,dashed,-] (m2) edge (m3);
    \draw [gray,dashed,-] (m1) edge[bend left=-30] (m3);
    \draw [gray,dashed,-] (y1) edge (y2);
    \draw [gray,dashed,-] (y2) edge (y3);
    \draw [gray,dashed,-] (y1) edge[bend left=30] (y3);
\end{tikzpicture}}
\end{minipage}
\begin{minipage}{.30\linewidth}
\subfloat[individual mediation effect]{
\begin{tikzpicture}[scale=0.55, transform shape]
    \node[state] (a) at (0,0) {$A_i$};
    \node[state] (m1) at (3,3) {$M_{i1}$};
    \node[state] (m2) at (3,0) {$M_{i2}$};
    \node[state] (m3) at (3,-3) {$M_{i3}$};
    \node[state] (y1) at (6,3) {$Y_{i1}$};
    \node[state] (y2) at (6,0) {$Y_{i2}$};
    \node[state] (y3) at (6,-3) {$Y_{i3}$};
    \draw (a) edge (m1);
    \draw (a) edge (m2);
    \draw (a) edge (m3);
    \draw (m1) edge (y1);
    \draw (m2) edge (y2);
    \draw (m3) edge (y3);
    \draw [gray,dashed,-] (m1) edge (m2);
    \draw [gray,dashed,-] (m2) edge (m3);
    \draw [gray,dashed,-] (m1) edge[bend left=-30] (m3);
    \draw [gray,dashed,-] (y1) edge (y2);
    \draw [gray,dashed,-] (y2) edge (y3);
    \draw [gray,dashed,-] (y1) edge[bend left=30] (y3);
\end{tikzpicture}}
\end{minipage}
\begin{minipage}{.30\linewidth}
\subfloat[spillover mediation effect]{
\begin{tikzpicture}[scale=0.55, transform shape]
    \node[state] (a) at (0,0) {$A_i$};
    \node[state] (m1) at (3,3) {$M_{i1}$};
    \node[state] (m2) at (3,0) {$M_{i2}$};
    \node[state] (m3) at (3,-3) {$M_{i3}$};
    \node[state] (y1) at (6,3) {$Y_{i1}$};
    \node[state] (y2) at (6,0) {$Y_{i2}$};
    \node[state] (y3) at (6,-3) {$Y_{i3}$};
    \draw (a) edge (m1);
    \draw (a) edge (m2);
    \draw (a) edge (m3);
    \draw (m1) edge (y2);
    \draw (m1) edge (y3);
    \draw (m2) edge (y1);
    \draw (m2) edge (y3);
    \draw (m3) edge (y1);
    \draw (m3) edge (y2);
    \draw [gray,dashed,-] (m1) edge (m2);
    \draw [gray,dashed,-] (m2) edge (m3);
    \draw [gray,dashed,-] (m1) edge[bend left=-30] (m3);
    \draw [gray,dashed,-] (y1) edge (y2);
    \draw [gray,dashed,-] (y2) edge (y3);
    \draw [gray,dashed,-] (y1) edge[bend left=30] (y3);
\end{tikzpicture}}
\end{minipage}
\caption{Causal graphs of the causal relationships among variables in a cluster with $N_i=3$ individuals. A dashed edge indicates generic association with unknown causal structure. We omit all pre-treatment variables, $\{N_i,\bcx_i,\bcv_i\}$, and their associated causal pathways, but acknowledge that all pre-treatment variables should have direct pathways towards all mediators and outcomes ($M_{ij}$ and $Y_{ij}$ for all $j=1,2,3$). {Panel (a) includes all pathways from treatment to the outcome. Panels (b)--(e) collect pathways associated with each mediation estimand.} 
\label{fig:dag}}
\end{figure}

\subsection{A motivating application: the \textit{Red de Protecci\'on Social} (RPS) trial}

Deficiencies in child nutrition remain a major public health concern in underdeveloped regions. The RPS trial (meaning ``Social Protection Network'' in Spanish) is a CRT designed to evaluate the effectiveness of a conditional cash transfer (CCT) program on households living in poverty across $K = 42$ \textit{comarcas} (administrative regions) in Nicaragua \citep{charters2023causal}. Randomization was conducted at the \textit{comarca} level, and only households in treated \textit{comarcas} received the conditional cash transfers. Our application aims to assess the role of household dietary diversity in mediating the treatment effect of the CCT program on child nutritional status. Specifically, the outcome is measured by child height-for-age $z$-scores, with higher scores indicating better nutritional status. This application extends the mediation analysis of \citet{charters2023causal}, who did not account for either the clustered design of CRTs or potential within-cluster interference. Given possible social interactions among households within clusters, we further examine how much of the indirect effect on child nutrition can be attributed to spillover influences from the dietary diversity of other households in the same cluster. Overall, our mediation framework provides deeper insight into causal pathways operating both within and across units in a cluster, thereby offering implications for improving policy design and implementation in CRTs.

\section{Causal estimands, assumptions, and identification}
\label{sec:mediation_effects}

For ease of presentation, we focus on $g(x,y)=x-y$, and extensions to ratio scales is discussed in Remark 1 in Supplementary Material. We first decompose $\text{TE}_C$ into a \textit{cluster-average natural indirect effect} ($\text{NIE}_C$) and a \textit{cluster-average natural direct effect} ($\text{NDE}_C$):
\begin{equation}\label{eq:te_decom1}
\text{TE}_C = \text{NIE}_C+\text{NDE}_C={g\left(\theta_C(1,1),\theta_C(1,0)\right)} + {g\left(\theta_C(1,0),\theta_C(0,0)\right)},
\end{equation}
with 
$
\theta_C(a,a^*) = \E\left[\frac{1}{N}\displaystyle\sum_{j=1}^{N} Y_{\cdot j}\left(a,M_{\cdot j}(a^*),\bcm_{\cdot (-j)}(a^*)\right)\right] \text{ for } a,a^*\in\{0,1\},
$
and $\theta_C(a,a)=\mu_C(a)$ by composition.
The $\text{NIE}_C$ defines a contrast between cluster-average potential outcomes under treatment, by switching the mediators in that cluster from their counterfactual values under control to factual values under treatment.  $\text{NDE}_C$  compares the cluster-average potential outcomes under different treatment conditions, but fixing the potential mediators in that cluster to their values under control. Intuitively, $\text{NDE}_C$ collects all direct causal pathways from treatment to outcome (Figure \ref{fig:dag}(b)) whereas $\text{NIE}_C$ collects the remaining indirect causal pathways that must involve the mediator (Figure \ref{fig:dag}(c)).  

To disentangle the role of each individual's mediator $M_{\cdot j}$ from that of other same-cluster members' mediator $\bcm_{\cdot (-j)}$, we can further decompose $\text{NIE}_C$ into a \textit{cluster-average spillover mediation effect} ($\text{SME}_C$) and a \textit{cluster-average individual mediation effect} ($\text{IME}_C$) as:
\begin{equation}\label{eq:te_decom2}
\text{NIE}_C = \text{SME}_C+\text{IME}_C={g\left(\theta_C(1,1),\tau_C\right)} + {g\left(\tau_C,\theta_C(1,0)\right)},
\end{equation}
where $\tau_C \!=\! \E\left[\frac{1}{N}\displaystyle\sum_{j=1}^{N} Y_{\cdot j}\left(1,M_{\cdot j}(1),\bcm_{\cdot (-j)}(0)\right)\right]$.  $\text{IME}_C$  compares cluster-average potential outcomes under treatment, by switching the individual's mediator from its value under control to that under treatment, but maintaining the mediators from all same-cluster members at their values without treatment. 
$\text{SME}_C$  compares cluster-average potential outcomes under treatment, by instead switching the remaining same-cluster members' mediators from their values under control to those under treatment, but maintaining an individual’s own mediator at its value under treatment. Thus, $\text{IME}_C$ investigates the indirect effect explained by each individual’s own mediator (illustrated by Figure \ref{fig:dag}(d)), whereas $\text{SME}_C$ tackles the spillover effect explained by mediators of other same-cluster individuals (illustrated by Figure \ref{fig:dag}(e)). Finally, the same decomposition applies to $\text{TE}_I=\text{NIE}_I+\text{NDE}_I={g\left(\theta_I(1,1),\theta_I(1,0)\right)} + {g\left(\theta_I(1,0),\theta_I(0,0)\right)}$ and $\text{NIE}_I=\text{SME}_I+\text{IME}_I={g\left(\theta_I(1,1),\tau_I\right)} + {g\left(\tau_I,\theta_I(1,0)\right)}$, where $\theta_I(a,a^*)$ and $\tau_I$ are defined in Remark 1 in Supplementary Material.

In the RPS trial, $\text{NIE}_C$ captures the effect of the CCT program on a child’s nutritional outcome within a household that operates through the dietary diversity conditions of all households in the comarca, whereas $\text{NDE}_C$ quantifies the direct effect of the CCT program not mediated by dietary diversity. The $\text{IME}_C$ estimand represents the effect of the CCT program that operates through a household's own dietary diversity to affect this household’s child nutrition, whereas $\text{SME}_C$ captures the effect of the CCT program through the dietary diversity of other households within the same comarca to affect a given household’s child nutritional status via unmeasured inter-household social interactions. By examining the $\text{IME}_C$ and $\text{SME}_C$, one can identify whether improvements in child nutrition are driven primarily by changes in a household’s own dietary behavior or by peer influences within the same community, thereby informing future refinement of interventions and scalability.


We focus on identification of $\theta_{V}(a,a^*)$ and $\tau_V$ for both $V\in\{C,I\}$, based on which all causal mediation effects are identified. 
We first state the identification assumptions.

\begin{assumption}\label{assum:consistency}
(\emph{Consistency}) $M_{ij}(a)=M_{ij}$ if $A_i=a$ and $Y_{ij}(a,m_{ij},\bmm_{i(-j)})=Y_{ij}$ if $A_i=a$ and $\{M_{ij},\bcm_{i(-j)}\}=\{m_{ij},\bmm_{i(-j)}\}$, for all $i$, $j$, $a\in\{0,1\}$ and $\{m_{ij},\bmm_{i(-j)}\}$.
\end{assumption}

\begin{assumption}\label{assum:randomization}
(\emph{Cluster randomization}) $A$ is randomized at the cluster level such that $A_i$ is independently drawn from a Bernoulli trial with $\Prob(A_i=1)=\pi \in(0,1)$. %
\end{assumption}

\begin{assumption}\label{assum:ignorability}
(\emph{Sequential ignorability})  $\{\bcm_i(1),\bcm_i(0)\} \!\perp \! Y_{ij}(a,\! m_{ij},\! \bmm_{i(-j)})|\{A_i, \allowdisplaybreaks \bcc_i, \allowdisplaybreaks N_i\}$, for all $i$, $j$, $a\in \{0,1\}$ and $\{m_{ij},\bmm_{i(-j)}\}$ over their valid support.
\end{assumption}

\begin{assumption}\label{assum:observed_data}
(\emph{Super-population sampling}) $\bco_1$,\dots,$\bco_K$ are mutually independent. For each cluster, $N$ follows a distribution $P_N$ over a finite support on $\mathbb{N}^{+}$. Conditional on $N$, the joint distribution $P_{\bcy,\bcm,A,\bcc|N}$ can be decomposed as $P_{\bcy|A,\bcm,\bcc,N}\times P_{\bcm|A,\bcc,N} \times P_{A} \times P_{\bcc|N}$ with each component having a finite second moment. Furthermore, positivity holds such that $f_{\bcm|A,\bcc,N}(\bmm|a,\bm c, n)>0$ for any $\{\bmm,a,\bm c, n\}$ over their valid support.
\end{assumption}
Assumption \ref{assum:consistency} allows for mediator interference within each cluster but rules out interference across clusters. Assumption \ref{assum:randomization} holds by design. Assumption \ref{assum:ignorability} extends the standard sequential ignorability assumption (\citealp{tchetgen2012semiparametric}) to clustered data, which is plausible if there are (i) no unmeasured baseline confounders for the mediator-outcome relationship and (ii) no treatment-induced confounders for the mediator-outcome relationship. In the RPS trial, Assumption \ref{assum:ignorability} holds if (i) we have collected sufficient baseline covariates that affect both household dietary and child nutrition (such as parental education, socioeconomic status, distance to markets) and (ii) no treatment-induced variables affect both household dietary and child nutrition. Note that Assumption \ref{assum:ignorability} still allows the existence of post-treatment variables but assumes such variables do not affect both mediator and outcome simultaneously (see Remark 2 in Supplementary Material for more discussions). For instance, child health check-up is measured after the CCT intervention and likely only affect child nutritional status. Assumption \ref{assum:ignorability} would be violated if health check-up also influences household dietary diversity. Assumption \ref{assum:observed_data} is a technical regularity condition for deriving asymptotic theory. 
Assumptions \ref{assum:consistency}--\ref{assum:observed_data} are sufficient for identifying $\theta_V(a,a^*)$ for $V\in\{C,I\}$. But to identify $\tau_V$, the following additional assumption is required.  
\begin{assumption}\label{assum:no_icc}
(\emph{Between-individual cross-world conditional independence}) 
Conditional on cluster size and all baseline covariates, $M_{ij}(1)\perp \bcm_{i(-j)}(0)|\{\bcc_i,N_i\}$ for all $i$ and $j$.
\end{assumption}
Assumption \ref{assum:no_icc} states that, after adjusting for $\bcc_i$ and $N_i$, an individual's potential mediator under treatment is independent from the potential mediators from all other individuals in the same cluster under control. Importantly, Assumption \ref{assum:no_icc} still allows for arbitrary residual dependence between single-world potential mediators within the same cluster ($M_{ij} (a)$ and $M_{ij'}(a)$) and intra-individual cross-world mediator dependence ($M_{ij} (a)$ and $M_{ij}(a^*)$ with $a\neq a^*$). In the RPS study, Assumption \ref{assum:no_icc} requires that, conditional on baseline covariates, the dietary diversity a household would exhibit under the CCP intervention is independent of the dietary diversity that other households within the same comarca would exhibit under the control condition. It is plausible when households vary considerably in socioeconomic status, health habits and dietary preferences, beyond which there exist no unmeasured factors that affect the counterfactual dietary diversity across different households.

To proceed with nonparametric identification, 
we introduce four nuisance functions of the observed data.  Specifically, define $\kappa(a,\bmm,\bm c,n)  =  f_{\bcm|A,\bcc,N}(\bmm|a,\bm c,n)$ as the joint density (probability) of mediators in a cluster conditional on assignment and covariates. Let $\kappa_{\cdot j}(a,m_{\cdot j},\bm c,n) = f_{M_{\cdot j}|A,\bcc,N}(m_{\cdot j}|a,\bm c,n)$ and $\kappa_{\cdot (-j)}(a,\bmm_{\cdot (-j)}, \bm c,n)=f_{\bcm_{\cdot (-j)}|A,\bcc,N}(\bmm_{\cdot (-j)}|a,\bm c,n)$ be the corresponding densities of $M_{\cdot j}$ and $\bcm_{\cdot (-j)}$. Notice that  $\kappa_{\cdot j}$ and $\kappa_{\cdot (-j)}$ can be derived from a joint mediator density with $\kappa_{\cdot j}(a,m_{\cdot j},\bm c,n)=\int_{\bmm_{\cdot (-j)}}\kappa(a,\bmm,\bm c,n) \differential \bmm_{\cdot (-j)}$ and $\kappa_{\cdot (-j)}(a,\bmm_{\cdot (-j)},\bm c,n)=\int_{m_{\cdot j}}\kappa(a,\bmm,\bm c,n) \differential m_{\cdot j}$. Define $\eta_{\cdot j}(a,\bmm,\bm c,n)  =  \E[Y_{\cdot j}|A=a,\bcm=\bmm,\bcc=\bm c,N=n]$ as the expectation of $Y_{\cdot j}$ conditional on  assignment, mediator, and covariates in that cluster. We abbreviate the nuisance functions as $h_{nuisance}^{(1)}=\{\eta_{\cdot j},\kappa,\kappa_{\cdot j},\kappa_{\cdot (-j)}\}$. For a quick reference, the list of all nuisance functions required are summarized in Table \ref{tab:definitions_of_nuisance}. 

\begin{theorem}\label{thm:identification}
Under Assumptions \ref{assum:consistency}--\ref{assum:observed_data}, we can identify
\begin{align*}
\theta_C(a,a^*) & = \E\left[\frac{1}{N}\sum_{j=1}^{N}\int_{\bmm} \eta_{\cdot j}(a,\bmm,\bcc,N)\kappa(a^*,\bmm,\bcc,N)  \differential \bmm \right],\\
\theta_I(a,a^*) & = \E\left[\sum_{j=1}^{N}\int_{\bmm} \eta_{\cdot j}(a,\bmm,\bcc,N)\kappa(a^*,\bmm,\bcc,N)\differential \bmm \right]\Big/\E[N],
\end{align*}
for any $a,a^*\in\{0,1\}$. 
Additionally, if Assumption \ref{assum:no_icc} holds, $\tau_C$ and $\tau_I$ can be identified by
\begin{align*}
\tau_C & = \E\left[\frac{1}{N}\sum_{j=1}^{N}\int_{\bmm} \eta_{\cdot j}(1,\bmm,\bcc,N)\kappa_{\cdot j}(1,m_{\cdot j},\bcc,N)\kappa_{\cdot (-j)}(0,\bmm_{\cdot (-j)},\bcc,N)\differential \bmm\right],\\
\tau_I & = \E\left[\sum_{j=1}^{N}\int_{\bmm} \eta_{\cdot j}(1,\bmm,\bcc,N)\kappa_{\cdot j}(1,m_{\cdot j},\bcc,N)\kappa_{\cdot (-j)}(0,\bmm_{\cdot (-j)},\bcc,N)\differential \bmm\right]\Big/\E[N].
\end{align*} 
\end{theorem}
Theorem \ref{thm:identification} generalizes the identification formulas in \cite{vanderweele2013mediation} to the scenario with informative cluster size. 
Based on Theorem \ref{thm:identification}, all mediation effects are identified as they are functions of $\theta_V(a,a^*)$ and $\tau_V$ ($V\in\{C,I\}$). Following \cite{tchetgen2012semiparametric}, we refer to identification formulas in Theorem \ref{thm:identification} as \textit{mediation functionals}. 

\begin{table}[!htbp]
\caption{A summary of definitions of nuisance functions and their requirement in constructing estimators based on the efficient influence functions.$^{ \mathparagraph \|}$}\label{tab:definitions_of_nuisance}
\centering
\vspace{-0.4cm}
\begin{threeparttable}
\scalebox{0.90}{
\begin{tabular}{clcccc}
\hline
Notation & Mathematical definition & $\!\!\!\!\!\!\widehat\theta_V^\text{eif$_1$}(a,a^*)\!\!\!\!$ & $\widehat\theta_V^\text{eif$_2$}(a,a^*)\!\!\!\!\!$ & $\widehat\tau_V^\text{eif$_1$}\!\!\!$ & $\!\!\!\!\!\widehat\tau_V^\text{eif$_2$}\!\!\!\!\!$ \\ \hline
 $\eta_{\cdot j}(a,\bmm,\bm c, n)$             &   $\E[Y_{\cdot j}|A=a,\bcm=\bmm,\bcc=\bm c,N=n]$         &  \checkmark  &  \checkmark & \checkmark & \checkmark\\ \hline
  $\eta_{\cdot j}^{\star}(a,a^*,\bm c,n)$             &    $\E\left[\eta_{\cdot j}(a,\bcm,\bcc, N) | A=a^*,\bcc= \bm c, N=n\right]$        &    &  \checkmark & & \\ \hline
  $\!\!\!\!\eta_{\cdot j}^{\dagger}(a,a^*,m_{\cdot j},\bm c,n)$ &   $\E[ \delta(\bcm,\!\bcc,\! N)|A\!=\!a^*,M_{\cdot j}\!=\!m_{\cdot j},\bcc\!=\!\bm c,N\!=\!n]^{\mathsection}$         &    &    & & \checkmark \\ \hline
  $\kappa(a,\bmm,\bm c, n)$                  &   $f_{\bcm|A,\bcc,N}(\bmm|a,\bm c,n)$         &  \checkmark  &    & \checkmark & \\ \hline
$\kappa_{\cdot j}(a,m_{\cdot j},\bm c, n)$                  &    $f_{M_{\cdot j}|A,\bcc,N}(m_{\cdot j}|a,\bm c,n)$        &    &   & \checkmark &  \checkmark \\ \hline
$\kappa_{\cdot j}^\star(a,\bmm,\bm c,n)$ &  $f_{M_{\cdot j}|\bcm_{\cdot (-j)},A,\bcc,N}(m_{\cdot j}|\bmm_{\cdot (-j)},a,\bm c, n)$          &    &    & & \checkmark \\ \hline
 $\!\!\!\kappa_{\cdot (\!-\!j)}(a,\!\bmm_{\cdot (-j)},\!\bm c, \!n)$                 &   $f_{\bcm_{\cdot (-j)}|A,\bcc,N}(\bmm_{\cdot j}|a,\bm c,n)$         &    &   & \checkmark & \\ \hline
 $s (a,\bmm,\bm c,n)$ &    $f_{A|\bcm,\bcc,N}(a|\bmm,\bm c,n)$        &    & \checkmark  & & \checkmark \\ \hline
\end{tabular}}
\begin{tablenotes}
      \footnotesize
      \item[$\mathparagraph$] For notational brevity,  the superscript `-par' or `-ml' in estimators are omitted. 
      \item[$\|$] Nuisance functions used in $\widehat \theta_V^{\text{mf-par}}(a,a^*)$ and $\widehat \tau_V^{\text{mf-par}}$ are identical to these used in $\widehat \theta_V^{\text{eif$_1$}}(a,a^*)$ and $\widehat \tau_V^{\text{eif$_1$}}$.
      \item[$\mathsection$] The function is defined as $\delta(\bcm,\bcc,N):=\eta_{\cdot j}(a,\bcm,\bcc, N){\kappa_{\cdot j}(a,M_{\cdot j},\bcc,N)}/{\kappa_{\cdot j}^\star(a^*,\bcm,\bcc,N)}$.
    \end{tablenotes}
\end{threeparttable}
\end{table}

\section{Semiparametric and data-adaptive nonparametric estimation}\label{sec:estimation}

\subsection{Specification of parametric working models}\label{sec:nuisance}


We first consider parametric models to estimate $h_{nuisance}^{(1)}$. 
Since $\{\kappa_{\cdot j},\kappa_{\cdot (-j)}\}$ can be specified from $\kappa$, only $\{\eta_{\cdot j},\kappa\}$ need to be modeled. For  $\eta_{\cdot j}(a,\bmm,\bm c,n)$, one can specify a mean model of $Y_{ij}$ conditional on
$A_i$, $\bcm_i$, $\bcc_i=\{\bcx_i,\bcv_i\}$, and $N_i$. 
Given that the dimensions of $\bcm_i$ and $\bcx_i$ can vary across clusters, a practical strategy is to adjust for summary functions with fixed dimensions in the regression model \citep{ogburn2022causal}. For example, a bivariate summary function $\left\{M_{ij},\frac{1}{N_i-1}\sum_{l=1,l\neq j}^{N_i}M_{il}\right\}$ of $\bcm_i$ can be considered such that $Y_{ij}$ is assumed to be affected by $\bcm_i$ via one's own mediator and the average mediator values of other same-cluster members. Similarly,  $\left\{\bcx_{ij},\frac{1}{N_i-1}\sum_{l=1,l\neq j}^{N_i}\bcx_{il}\right\}$ can be used for $\bcx_i$.

Modeling $\kappa(a,\bmm,\bm c,n)$ requires specification of a joint density of mediators within the same cluster. 
We define $\kappa(a,\bmm,\bm c,n)$ with two variationally independent components: the marginal mediator probability for each individual $\kappa_{\cdot j}(a,m_{\cdot j},\bm c,n)$ and a copula $\mathcal C$ characterizing the association structure.  
By the Sklar's theorem, there exists a copula $\mathcal C$ such that 
\begin{equation}\label{eq:CDF_copula}
P_{\bcm|A,\bcc,N}(\bmm|a,\bm c,n) = \mathcal C(\mathcal K_{\cdot 1}(m_{\cdot 1}),\cdots,\mathcal K_{\cdot n}(m_{\cdot n})|a,\bm c,n),
\end{equation}
where $\mathcal K_{\cdot j}(m_{\cdot j}):=\int_{-\infty}^{m_{\cdot j}}\mathcal \kappa_{\cdot j}(a,t,\bm c, n) \differential t$ is the CDF of $M_{\cdot j}$ and $\mathcal C(u_{\cdot 1},\dots,u_{\cdot n}|a,\bm c,n)$ is a $n$-variate copula supported on $\{u_{\cdot 1},\cdots,u_{\cdot n}\}\in [0,1]^n$.  
When the mediator is continuous,
 we have $\kappa(a,\bmm,\bm c,n)=\kappa^{c}(a,\bmm,\bm c,n)  := c(\mathcal K_{\cdot 1}(m_{\cdot 1}),\cdots,\mathcal K_{\cdot n}(m_{\cdot n})|a,\bm c,n) \prod_{j=1}^n \kappa_{j}(a,m_{\cdot j},\bm c,n)$, where $c(u_{\cdot 1},\dots,u_{\cdot n}|a,\bm c,n)=\frac{\partial^n}{\partial u_{\cdot 1}\cdots \partial u_{\cdot n}}\mathcal C (u_{\cdot 1},\dots,u_{\cdot n}|a,\bm c,n)$ is the density of the copula $\mathcal C$. Expression of $\kappa(a,\bmm,\bm c,n)$ with a discrete mediator is given in Web Appendix A.1.

To proceed, one can specify a working regression model (e.g. generalized linear model) to obtain $\widehat\kappa_{\cdot j}(a,m_{\cdot j},\bm c,n)$, 
along with a parametric multivariate copula as in \eqref{eq:CDF_copula}. We focus on a parametric Gaussian copula  model \citep{masarotto2012gaussian} to describe  $\mathcal C$, which leverages a $n$-by-$n$ correlation matrix with unknown parameter $\bm \rho$ to describe the dependence structure for the mediators across individuals in the same cluster. Under the Gaussian copula and given $\widehat\kappa_{\cdot j}(a,m_{\cdot j},\bm c,n)$,  $\bm\rho$  can be estimated by the pseudo-likelihood approach outlined in Web Appendix A.1; estimates of $\{\kappa,\kappa_{\cdot (-j)}\}$ are then obtained accordingly and details are given in Web Appendix A.1. To summarize, $h_{nuisance}^{(1)}=\{\eta_{\cdot j}, \kappa,\kappa_{\cdot j},\kappa_{\cdot (-j)}\}$ can be estimated by $\widehat h_{nuisance}^{(1)}=\{\widehat\eta_{\cdot j}, \widehat\kappa^c,\widehat\kappa_{\cdot j},\widehat \kappa_{\cdot (-j)}^c\}$ based on three parametric models of $\eta_{\cdot j}(a,\bmm,\bm c,n)$, $\kappa_{\cdot j}(a,m_{\cdot j},\bm c,n)$, and $\mathcal C(u_{\cdot 1},\dots,u_{\cdot n}|a,\bm c,n)$.

Based on Theorem \ref{thm:identification}, $\theta_C(a,a^*)$ and $\tau_C$ are estimated by $\widehat\theta_C^{\text{mf-par}}(a,a^*)=\frac{1}{K}\sum_{i=1}^K\frac{1}{N_i}\sum_{j=1}^{N_i}\mathcal I_{1,ij}$ and $\widehat\tau_C^{\text{mf-par}} = \frac{1}{K}\sum_{i=1}^K\frac{1}{N_i}\sum_{j=1}^{N_i}\mathcal I_{2,ij}$, where $\mathcal I_{1,ij} = \int_{\bmm_i} \widehat\eta_{ij}(a,\bmm_i,\bcc_i,N_i)\widehat\kappa^c(a^*,\bmm_i,\bcc_i,N_i)  \differential \bmm_i$ and 
$\mathcal I_{2,ij}  = \int_{\bmm_i} \widehat \eta_{i j}(1,\bmm_i,\bcc_i,N_i)\widehat \kappa_{ij}(1,m_{ij},  \bcc_i,N_i) \widehat \kappa_{i (-j)}^c(0,\bmm_{i (-j)},\bcc_i,N_i)\differential \bmm_i$. Analogously, we have $\widehat\theta_I^{\text{mf-par}}(a,a^*)=\frac{1}{K\times\overline N}\displaystyle\sum_{i=1}^K\sum_{j=1}^{N_i}\mathcal I_{1,ij}$ and $\widehat\tau_I^{\text{mf-par}} = \frac{1}{K\times\overline N}\displaystyle\sum_{i=1}^K\sum_{j=1}^{N_i}\mathcal I_{2,ij}$, where $\overline N =\frac{1}{K}\sum_{i=1}^K N_i$ is the average cluster size. The integrals, $\mathcal I_{1,ij}$ and $\mathcal I_{2,ij}$, can be calculated through Monte Carlo integration (see the end of Web Appendix A.1 of Supplementary Material).

\subsection{Semiparametric doubly robust estimators}
\label{sec:DR_estimators}

To improve upon the fully parametric estimators via the mediation functionals, we leverage the theory of semiparametric inference \citep{bickel1993efficient} to develop more robust estimators. We first derive the EIF of each mediation estimand under a nonparametric model, where the observed data likelihood is left unrestricted (except for the known treatment assignment). 
To introduce the EIFs, we define the following six auxiliary functions of $h_{nuisance}^{(1)}$:
\begin{align}
& w^{(1)}(a,a^*,\bmm,\bm c,n)  = \frac{\kappa (a^*,\bmm,\bm c,n)}{\kappa (a,\bmm,\bm c,n)}, w^{(2)}(a,a^*,a',\bmm,\bm c,n)  = 
\frac{\kappa_{\cdot j} (a,m_{\cdot j},\bm c,n)\kappa_{\cdot (-j)} (a^*,\bmm_{\cdot (-j)},\bm c,n)}{\kappa (a',\bmm,\bm c,n)},\nonumber \\
& u_{\cdot j}^{(1)}(a,a^*,\bm c,n)  =  \int_{\bmm} \eta_{\cdot j}(a,\bmm,\bm c, n) \kappa (a^*,\bmm, \bm c, n) \differential \bmm , \nonumber\\
& u_{\cdot j}^{(2)}(a,a^*,m_j,\bm c,n)  =  \int_{\bmm_{\cdot (-j)}} \eta_{\cdot j}(a,\bmm,\bm c, n) \kappa_{\cdot (-j)} (a^*,\bmm_{\cdot(-j)}, \bm c, n) \differential \bmm_{\cdot (-j)},\label{eq:aux_functions}\\
& u_{\cdot j}^{(3)}(a,a^*,\bmm_{\cdot (-j)},\bm c,n)  =  \int_{m_{\cdot j}} \eta_{\cdot j}(a,\bmm,\bm c, n) \kappa_{\cdot j} (a^*,m_j, \bm c, n) \differential m_{\cdot j}, \nonumber\\
& u_{\cdot j}^{(4)}(a,a^*,a',\bm c,n)   =  \int_{\bmm} \eta_{\cdot j}(a,\bmm,\bm c, n) \kappa_{\cdot j} (a^*,m_{\cdot j}, \bm c, n) \kappa_{\cdot (-j)} (a',\bmm_{\cdot(-j)}, \bm c, n) \differential \bmm, \nonumber
\end{align}
where $\{w^{(1)},w^{(2)}\}$ are ratios of mediator densities and $\{u_{\cdot j}^{(1)},u_{\cdot j}^{(2)},u_{\cdot j}^{(3)},u_{\cdot j}^{(4)}\}$ are integrals of $\eta_{\cdot j}(a,\bmm,\bm c, n)$ with respect to distribution of the mediators under different assignments. 
\begin{theorem}\label{thm:eif}
 Under Assumptions \ref{assum:consistency}--\ref{assum:observed_data},  
 the EIFs of $\theta_C(a,a^*)$ and $\theta_I(a,a^*)$ are $\mathcal{D}_{\theta_C(a,a^*)}(\bco) = \psi_{\theta}(a,a^*;\bco) \!-\! \theta_C(a,a^*)$ and $\mathcal{D}_{\theta_I(a,a^*)}(\bco) = \frac{N}{\E[N]}\left\{\psi_{\theta}(a,a^*;\bco) \!-\! \theta_I(a,a^*)\right\}$, 
respectively, where
\begin{align*}
\psi_{\theta}(a,a^*;\bco) = &  \frac{1}{N}\sum_{j=1}^N \Bigg\{\frac{\mathbb{I}(A=a)}{\pi^a(1-\pi)^{1-a}} w^{(1)}(a,a^*,\bcm,\bcc,N) \left\{Y_{\cdot j}-\eta_{\cdot j}(a,\bcm,\bcc,N)\right\} \\
& +  \frac{\mathbb{I}(A=a^*)}{\pi^{a^*}(1-\pi)^{1-a^*}}\left\{\eta_{\cdot j}(a,\bcm,\bcc,N) - u_{\cdot j}^{(1)}(a,a^*,\bcc,N)\right\} + u_{\cdot j}^{(1)}(a,a^*,\bcc,N) \Bigg\}.
\end{align*}
Additionally, if Assumption \ref{assum:no_icc} holds, the EIFs of $\tau_C$ and $\tau_I$ are $\mathcal{D}_{\tau_C}(\bco) = \psi_{\tau}(\bco) - \tau_C$ and $\mathcal{D}_{\tau_I}(\bco) = \frac{N}{\E[N]}\left\{\psi_{\tau}(\bco) - \tau_I\right\}$,  respectively, where
\begin{align*}
\psi_{\tau}(\bco) = & \frac{1}{N}\sum_{j=1}^N \Bigg\{\frac{A}{\pi} w^{(2)}(1,0,1,\bcm,\bcc,N)\left\{Y_{\cdot j}-\eta_{\cdot j}(1,\bcm,\bcc,N)\right\} \\
& + \frac{A}{\pi} \left\{u_{\cdot j}^{(2)}(1,0,M_{\cdot j},\bcc,N) - u_{\cdot j}^{(4)}(1,1,0,\bm C,N) \right\} \\
& + \frac{1-A}{1-\pi} \left\{u_{\cdot j}^{(3)}(1,1,\bcm_{\cdot (-j)},\bcc,N) - u_{\cdot j}^{(4)}(1,1,0,\bm C,N) \right\} + u_{\cdot j}^{(4)}(1,1,0,\bm C,N) \Bigg\}.
\end{align*}
Therefore, the semiparametric efficiency lower bound for each causal estimand is $\E[\{\mathcal{D}_{\zeta}(\bco)\}^2]$, for $\zeta\in\{\theta_C(a,a^*), \theta_I(a,a^*), \tau_C,\tau_I\}$.
\end{theorem}


Theorem \ref{thm:eif} generalizes of the EIFs of mediation functionals from the independent data setting \citep{tchetgen2012semiparametric} to CRTs; see Remark 3 in Supplementary Material for additional discussion. 
Theorem \ref{thm:eif} shows that EIFs for the individual-average mediation functionals bear a similar pattern to their cluster-average counterparts, except that the former additionally include a cluster size multiplier, ${N}/{\E[N]}$, to target the population of all individual units. Importantly, the EIFs are functions of $h_{nuisance}^{(1)}$ through the six auxiliary functions $\{w^{(1)},w^{(2)},u_{\cdot j}^{(1)},u_{\cdot j}^{(2)},u_{\cdot j}^{(3)},u_{\cdot j}^{(4)}\}$ defined in \eqref{eq:aux_functions}, and directly motivate new causal mediation estimators that optimally combine information across all working models. 
Based on working models in Section \ref{sec:nuisance}, we propose the following semiparametric estimators:
\begin{align}
\widehat{\theta}_C^{\text{eif$_1$-par}}(a,a^*) = \frac{1}{K}\sum_{i=1}^K \widehat\psi_{\theta}(a,a^*;\bco_i), & \quad \widehat\tau_C^{\text{eif$_1$-par}} = \frac{1}{K}\sum_{i=1}^K \widehat\psi_\tau(\bco_i), \nonumber \\
\widehat{\theta}_I^{\text{eif$_1$-par}}(a,a^*) = \frac{1}{K \overline N}\sum_{i=1}^K N_i\times \widehat\psi_{\theta}(a,a^*;\bco_i) , & \quad \widehat\tau_I^{\text{eif$_1$-par}} = \frac{1}{K \overline N}\sum_{i=1}^K N_i\times \widehat\psi_\tau(\bco_i),\label{eq:estimator_eif1}
\end{align}
where $\{\widehat\psi_{\theta}(a,a^*;\bco_i),\widehat\psi_\tau(\bco_i)\}$ are plugin estimates of $\{\psi_{\theta}(a,a^*;\bco_i),\psi_\tau(\bco_i)\}$ based on $\widehat h_{nuisance}^{(1)}$. Evaluation of $\{\widehat u_{\cdot j}^{(1)},\widehat u_{\cdot j}^{(2)},\widehat u_{\cdot j}^{(3)},\widehat u_{\cdot j}^{(2)}\}$ in $\{\widehat\psi_{\theta}(a,a^*\!;\!\bco_i),\widehat\psi_\tau(\bco_i)\}$ requires calculating multivariate integrals, and can be obtained from Monte Carlo integration.



While semiparametric estimators \eqref{eq:estimator_eif1} directly follow from the EIF, they 
necessitate the calculation of multi-dimensional integrals with respect to the mediator density functions. To address this  limitation, we next reparameterize the six auxiliary functions in \eqref{eq:aux_functions}.



\begin{proposition}\label{prop:reparametrization} Define two densities $s (a,\bmm,\bm c,n) = f_{A|\bcm,\bcc,N}(a|\bmm,\bm c,n)$, $\kappa_{\cdot j}^\star(a,\bmm,\bm c,n) = f_{M_{\cdot j}|\bcm_{\cdot (-j)},A,\bcc,N}(m_{\cdot j}|\bmm_{\cdot (-j)},a,\bm c, n)$, and two expectations
\begin{align*}
\eta_{\cdot j}^{\star}(a,a^*,\bm c,n)  & =  \E\left[\eta_{\cdot j}(a,\bcm,\bcc, N) | A=a^*,\bcc= \bm c, N=n\right], \\
\eta_{\cdot j}^{\dagger}(a,a^*,m_{\cdot j},\bm c,n) & =  \E\left[ \eta_{\cdot j}(a,\bcm,\bcc, N)\frac{\kappa_{\cdot j}(a,M_{\cdot j},\bcc,N)}{\kappa_{\cdot j}^\star(a^*,\bcm,\bcc,N)}\Big|A\!=\!a^*,M_{\cdot j}\!=\!m_{\cdot j},\bcc\!=\!\bm c,N\!=\!n\right].
\end{align*}
Then, the auxiliary functions in \eqref{eq:aux_functions} can be re-expressed as:
\begin{align*}
 w^{(1)}(a,a^*,\bmm,\bm c,n) & = \frac{s (a^*,\bmm,\bm c,n)}{s (a,\bmm,\bm c,n)}\times \frac{\pi^{a}(1-\pi)^{1-a}}{\pi^{a^*}(1-\pi)^{1-a^*}}, \\
 w^{(2)}(a,a^*,a',\bmm,\bm c,n) & = 
\frac{\kappa_{\cdot j}(a,m_{\cdot j},\bm c,n)}{\kappa_{\cdot j}^\star(a^*,\bmm,\bm c,n)} \times \frac{s (a^*,\bmm,\bm c,n)}{s (a',\bmm,\bm c,n)}\times \frac{\pi^{a'}(1-\pi)^{1-a'}}{\pi^{a^*}(1-\pi)^{1-a^*}},
\end{align*}
$u_{\cdot j}^{(1)}(a,a^*,\bm c,n) =  \eta_{\cdot j}^{\star}(a,a^*,\bm c,n)$, $u_{\cdot j}^{(2)}(a,a^*,m_j,\bm c,n) =  \eta_{\cdot j}^{\dagger}(a, a^*, m_j, \bm c, n)$, $u_{\cdot j}^{(3)}(a, a^*, \bmm_{\cdot (\!-\!j)},\! \bm c,\! n)$ is defined in \eqref{eq:aux_functions}, and $u_{\cdot j}^{(4)}(a,a^*,a',\bm c,n) =  \int_{m_{\cdot j}} \eta_{\cdot j}^{\dagger}(a,a',m_j,\bm c,n)  \kappa_{\cdot j}(a^*,m_{\cdot j}, \bm c, n) \differential m_{\cdot j}$. 
\end{proposition}
{Similar low-dimensional reparameterization has been considered in previous semiparametric mediation analysis framework for handling post-treatment confounders \citep{diaz2021nonparametric} or multiple mediators \citep{zhou2022semiparametric}.} Proposition \ref{prop:reparametrization} permits us to use nuisance $h_{nuisance}^{(2)}\!=\!\{\eta_{\cdot j}(a,\bmm,\bm c,n), \! \eta_{\cdot j}^{\star}(a,a^*,\bm c,n),\! \eta_{\cdot j}^{\dagger}(a,a^*,m_{\cdot j},\bm c,n),\! \kappa_{\cdot j}(a,\bmm,\bm c,n),\! \allowbreak \kappa_{\cdot j}^\star(a,\bmm,\bm c,n), \! s (a,\bmm,\bm c,n)\}$, where  $\{\eta_{\cdot j},\kappa_{\cdot j}\}$ are recycled from $h_{nuisance}^{(1)}$ but $\{\eta_{\cdot j}^{\star},\eta_{\cdot j}^{\dagger},\kappa_{\cdot j}^\star,s\}$ are new lower-dimensional nuisance functions after reparameterization (see Table \ref{tab:definitions_of_nuisance} for definitions of all nuisance functions). 

Proposition \ref{prop:reparametrization}  plays an important role in operationalizing the semiparametric estimators based on the EIFs. 
First, the EIFs under reparameterization only include one-dimensional integrals (rather than multi-dimensional ones), which can be conveniently obtained from numerical integration routines. Second, all components in $h_{nuisance}^{(2)}$ are one-dimensional conditional density or expectations; this enables the specification of familiar statistical models, and dispenses with the specification of an association model for a multivariate density function. 
For example, $s(a,\bmm,\bm c,n)$ can be estimated by regressing $A_i$ on $\bm V_i$, $N_i$, and cluster-level summary functions of $\bcx_i$ and $\bcm_i$. Estimation of $\kappa_{\cdot j}^\star(a,\bmm,\bm c,n)$ is similar to $\kappa_{\cdot j}(a,m_{\cdot j},\bm c,n)$, except for the adjustment of a summary function of $\bcm_{i(-j)}$. Finally, $\widehat \eta_{\cdot j}^{\star}(a,a^*,\bm c,n)$ can be obtained by regressing $\widehat \eta_{ij}(a,\bcm_i,\bcc_i,N_i)$ on $A$, $\bcv_i$, $N_i$ and a summary function of $\bcx_i$, and $\widehat\eta_{\cdot j}^{\dagger}(a,a^*,m_{\cdot j},\bm c,n)$ can be obtained by regressing $\widehat\eta_{i j}(a,\bcm_i,\bcc_i, N_i)\frac{\widehat \kappa_{ij}(a,M_{ij},\bcc_i,N_i)}{\widehat \kappa_{ij}^\star(a^*,\bcm_i,\bcc_i,N_i)}$ on $A_i$, $M_{ij}$, $\bcv_i$, $N_i$, and a summary function of $\bcx_i$. We offer practical strategies in Web Appendix A.2 for specifying parametric models of $h_{nuisance}^{(2)}$ that ensures approximate compatibility.

Proposition \ref{prop:reparametrization} motivates alternative semiparametric estimators; we denote the new estimators by $\widehat \zeta^{\text{eif$_2$-par}}$, for $\zeta\in\{\theta_C(a,a^*),\theta_I(a,a^*),\tau_C,\tau_I\}$. The estimators have the same formulation as \eqref{eq:estimator_eif1}, but  $\widehat \psi_{\theta}(a,a^*,\bm O)$ and $\widehat \psi_{\tau}(\bm O)$ are obtained based on $\widehat h_{nuisance}^{(2)}$. Theorem \ref{thm:parametric_theory} summarizes asymptotic properties of the two sets of semiparametric estimators, $\widehat \zeta^{\text{eif$_1$-par}}$ and $\widehat \zeta^{\text{eif$_2$-par}}$.

\begin{theorem}\label{thm:parametric_theory}
Suppose that the nuisance functions are estimated via parametric working models. 
The following results hold for $\theta_V(a,a^*)$ and $\tau_V$ with both $V\in\{I,C\}$, and expressions of asymptotic variances are presented in the Supplementary Material.
\begin{compactitem}
\item[(i)] If  $\{\kappa_{\cdot j},\mathcal C\}$ or $\eta_{\cdot j}$ is correctly specified,  
$\sqrt{K}\left\{\widehat\theta_{V}^{\text{eif$_1$-par}}(a,a^*)-\theta_{V}(a,a^*)\right\}\overset{d}{\to} N(0,\Sigma_{\theta_V(a,a^*)}^{\text{eif$_1$-par}})$; if $\{\eta_{\cdot j},\kappa_{\cdot j},\mathcal C\}$ are correctly specified, $\Sigma_{\theta_V(a,a^*)}^{\text{eif$_1$-par}}=\E[\mathcal D_{\theta_{V}(a,a^*)}(\bco)^2]$ achieves the efficiency lower bound for estimating $\theta_{V}(a,a^*)$. If  $\{\kappa_{\cdot j},\mathcal C\}$ or $\{\kappa_{\cdot j},\eta_{\cdot j}\}$ are correctly specified, then $\sqrt{K}(\widehat\tau_{V}^{\text{eif$_1$-par}}-\tau_{V})\overset{d}{\to} N(0,\Sigma_{\tau_V}^{\text{eif$_1$-par}})$; if $\{\eta_{\cdot j},\kappa_{\cdot j},\mathcal C\}$ are correctly specified, $\Sigma_{\tau_V}^{\text{eif$_1$-par}}=\E[\mathcal D_{\tau_{V}}(\bco)^2]$ achieves the efficiency lower bound for estimating $\tau_{V}$.

\item[(ii)] If  $\eta_{\cdot j}$ or $s$ is correctly specified, then $\sqrt{K}\left\{\widehat\theta_{V}^{\text{eif$_2$-par}}(a,a^*)-\theta_{V}(a,a^*)\right\}\overset{d}{\to} N(0,\Sigma_{\theta_V(a,a^*)}^{\text{eif$_2$-par}})$; if $\{\eta_{\cdot j},\eta_{\cdot j}^{\star},s\}$ are correctly specified, $\Sigma_{\theta_V(a,a^*)}^{\text{eif$_2$-par}}=\E[\mathcal D_{\theta_{V}(a,a^*)}(\bco)^2]$ achieves the efficiency lower bound for estimating $\theta_{V}(a,a^*)$. If $\{\kappa_{\cdot j},\eta_{\cdot j}\}$ or $\{\kappa_{\cdot j},\kappa_{\cdot j}^{\star},s\}$ are correctly specified, then  $\sqrt{K}(\widehat\tau_{V}^{\text{eif$_2$-par}}\\ -\tau_{V})\overset{d}{\to} N(0,\Sigma_{\tau_V}^{\text{eif$_2$-par}})$; if $\{\eta_{\cdot j},\eta_{\cdot j}^{\dagger},\kappa_{\cdot j},\kappa_{\cdot j}^{\star},s\}$ are correctly specified, $\Sigma_{\tau_V}^{\text{eif$_2$-par}}=\E[\mathcal D_{\tau_{V}}(\bco)^2]$ achieves the efficiency lower bound for estimating $\tau_{V}$.
\end{compactitem}
\end{theorem}
Theorem \ref{thm:parametric_theory} indicates that the two sets of semiparametric estimators of $\theta_V(a,a^*)$ (i.e., $\widehat \theta_V^{\text{eif$_1$-par}}(a,a^*)$ and $\widehat \theta_V^{\text{eif$_2$-par}}(a,a^*)$) are doubly robust as they are $\sqrt{K}$-consistent and asymptotically normal (CAN) when one of two separate sets of parametric working models are correctly specified. In contrast to $\theta_V(a,a^*)$, the two semiparametric estimators of $\tau_V$ are \emph{conditional} doubly robust requiring that $\kappa_{\cdot j}$ is correctly specified. Specifically, when $\kappa_{\cdot j}$ is correct, $\widehat\tau_V^{\text{eif$_1$-par}}$ is CAN if $\mathcal C$ or $\eta_{\cdot j}$ is correctly specified whereas $\widehat\tau_V^{\text{eif$_2$-par}}$ is CAN if $\eta_{\cdot j}$ or $\{\kappa_{\cdot j}^\star,s\}$ are correctly specified. Interestingly, the consistency of $\{\widehat \theta_V^{\text{eif$_2$-par}}(a,a^*),\widehat \tau_V^{\text{eif$_2$-par}}\}$ does not depend on the specification of $\{\eta_{\cdot j}^{\star},\eta_{\cdot j}^{\dagger}\}$, but their local efficiency necessitates the correct specification of $\{\eta_{\cdot j}^{\star},\eta_{\cdot j}^{\dagger}\}$. Remark 4 in Supplementary Material clarifies additional ``bonus'' robustness properties of our estimators when $\eta_{\cdot j}$ follows a linear regression without mediator-by-covariate interactions.

Finally, since each asymptotic variance expression in Theorem \ref{thm:parametric_theory} has a rather complicated form, we recommend nonparametric cluster bootstrap \citep{field2007bootstrapping} for inference, which provides consistent confidence intervals (see Remark 5 in Supplementary Material). 

\subsection{Leveraging machine learning to obtain fully efficient estimators}
\label{sec:ML_estimators}

We extend $\widehat\zeta^{\text{eif$_1$-par}}$ and $\widehat\zeta^{\text{eif$_2$-par}}$ (for $\zeta\in\{\theta_C(a,a^*),\theta_I(a,a^*),\tau_C,\tau_I\}$) by leveraging  machine learners to estimate the nuisance functions. This leads to two machine learning estimators $\widehat\zeta^{\text{eif$_1$-ml}}$ and $\widehat\zeta^{\text{eif$_2$-ml}}$. {As $h_{nusiance}^{(2)}$ only involves conditional expectations or conditional densities of univariate variables, several off-the-shelf machine learners can be employed to obtain their estimates (see \cite{phillips2023practical} for a list of machine learners and a guide on optimizing their performance through the Super Learner).} Using machine learners to estimate $h_{nusiance}^{(1)}$ may be more challenging, because $h_{nusiance}^{(1)}$ involves two multi-dimensional conditional densities $\{\kappa,\kappa_{\cdot (-j)}\}$. Similar to Section \ref{sec:nuisance}, we parameterize $h_{nusiance}^{(1)}$ into $\{\eta_{\cdot j},\kappa^c,\kappa_{\cdot j},\kappa_{\cdot (-j)}^c\}$ based on a multivariate copula $\mathcal C$ given in \eqref{eq:CDF_copula}. Then, $\widehat\eta_{\cdot j}$ and $\widehat\kappa_{\cdot j}$ can still be based on machine learners designed for conditional expectation and conditional density. For modeling $\mathcal C$, we also consider the Gaussian copula, where the unknown association parameter $\widehat{\bm \rho}$ is obtained by the pseudo-likelihood approach except that the machine learning estimator $\widehat \kappa_{\cdot j}$ is plugged in the pseudo-likelihood (Web Appendix A.1). 
When machine learners are applied, cross-fitting is necessary to control the empirical process term to be asymptotically negligible \citep{chernozhukov2018double}, and details are provided in Web Appendix A.3. 
The asymptotic properties (including the rate robustness results) of $\widehat\zeta^{\text{eif$_1$-ml}}$ and $\widehat\zeta^{\text{eif$_2$-ml}}$ are provided in Theorem 4 in Web Appendix A.3. The rate robustness properties of $\widehat\zeta^{\text{eif$_1$-ml}}$ and $\widehat\zeta^{\text{eif$_2$-ml}}$ echo their doubly robust counterparts when all nuisances are specified by parametric models. Furthermore, Theorem 4 in Web Appendix A.3 also reveals that $\widehat\zeta^{\text{eif$_1$-ml}}$ and $\widehat\zeta^{\text{eif$_2$-ml}}$ are also $\sqrt{K}$-consistent, asymptotically normal and semiparametrically efficient when all nuisance functions involved in each estimator are consistent in $L_2(P)$-norm and satisfy mild conditions for convergence rate. A $o_p(K^{-1/4})$-type convergence rate among nuisance functions is sufficient to ensure all $\widehat\zeta^{\text{eif$_1$-ml}}$ and $\widehat\zeta^{\text{eif$_2$-ml}}$ to be consistent, asymptotically normal, and efficient. 
For inference, we use the empirical variance of EIF under cross-fitting, where $\text{Var}(\widehat\tau_C^{\text{eif$_1$-ml}})$ and $\text{Var}(\widehat\tau_I^{\text{eif$_1$-ml}})$ can be estimated by $\frac{1}{K^2}\sum_{i=1}^K\left\{\widehat \psi_\tau(\bco_i)-\widehat\tau_C^{\text{eif$_1$-ml}}\right\}^2$ and $\frac{1}{K^2}\sum_{i=1}^K\left\{\frac{N_i}{\overline{N}}[\widehat \psi_\tau(\bco_i)-\widehat\tau_I^{\text{eif$_1$-ml}}]\right\}^2$, respectively.

\subsection{Stabilization to improve finite-sample performance}\label{sec:stablization}

All EIF-based estimators involve density ratios, $w^{(1)}$ and $w^{(2)}$, regardless of whether the reparameterization in Proposition \ref{prop:reparametrization} is used. If $\widehat w^{(1)}$ and $\widehat w^{(2)}$ are highly variable, then the weighting estimators may be less stable. In Web Appendix A.4, we develop a stabilization procedure to improve the stability of our estimators, in a similar spirit to the targeted minimum-loss based estimation \citep{van2011targeted}. In Web Appendix A.4, we show that stabilzation does not change the asymptotics of the original, non-stabilized estimators.

In Table \ref{tab:estimator_summary}, we summarize all  estimators that leverages the EIFs to estimate $\theta_V(a,a^*)$ and $\tau_V$ (with $V\in\{C,I\}$). This includes a total of 8 estimators, distinguished by three key factors: whether the original or  reparameterized EIF are leveraged, whether parametric working models or machine learners are used for estimating the nuisance functions, and whether the stabilization procedure is applied. {
For completeness, the computational details for  the mediation effect estimands are given in Web Appendix A.5.}

\begin{table}[]
\caption{A summary of the proposed semiparametric estimators of $\theta_V(a,a^*)$ and $\tau_V$, with $V\in\{C,I\}$, and the sections where the estimators have been defined. To make the notation more explicit, we add the superscript `-ns' to represent non-stabilized estimators and the superscript `-s' to represent stabilized estimators. The asymptotic properties between the non-stabilized estimators and stabilized estimators are identical. \label{tab:estimator_summary}}
\centering
\vspace{-0.4cm}
\scalebox{0.8}{
\begin{tabular}{cc|cc|cc}
\hline
\multirow{2}{*}{Stabilization}                   & \multirow{2}{*}{Models}           & \multicolumn{2}{c|}{Original EIF for} & \multicolumn{2}{c}{Reparameterized EIF for } \\ 
& & \multicolumn{2}{c|}{$\theta_V(a,a^*)$ and $\tau_V$ estimands} & \multicolumn{2}{c}{$\theta_V(a,a^*)$ and $\tau_V$ estimands} \\
\hline
\multirow{4}{*}{Non-stabilized} & \multirow{2}{*}{Parametric}      &    $\widehat\theta_V^{\text{eif$_1$-par-ns}}(a,a^*)$ & $\widehat\tau_V^{\text{eif$_1$-par-ns}}$          &        $\widehat\theta_V^{\text{eif$_2$-par-ns}}(a,a^*)$ & $\widehat\tau_V^{\text{eif$_2$-par-ns}}$             \\ 
&  & Section \ref{sec:DR_estimators} & Section \ref{sec:DR_estimators} & Section \ref{sec:DR_estimators} & Section \ref{sec:DR_estimators} \\
\cline{2-6} 
& \multirow{2}{*}{Machine Learner}  &     $\widehat\theta_V^{\text{eif$_1$-ml-ns}}(a,a^*)$ & $\widehat\tau_V^{\text{eif$_1$-ml-ns}}$          &        $\widehat\theta_V^{\text{eif$_2$-ml-ns}}(a,a^*)$ & $\widehat\tau_V^{\text{eif$_2$-ml-ns}}$             \\
&  & Section \ref{sec:ML_estimators} & Section \ref{sec:ML_estimators} & Section \ref{sec:ML_estimators} & Section \ref{sec:ML_estimators} \\
                               \hline
\multirow{4}{*}{Stabilized} & \multirow{2}{*}{Parametric}      &    $\widehat\theta_V^{\text{eif$_1$-par-s}}(a,a^*)$ & $\widehat\tau_V^{\text{eif$_1$-par-s}}$          &        $\widehat\theta_V^{\text{eif$_2$-par-s}}(a,a^*)$ & $\widehat\tau_V^{\text{eif$_2$-par-s}}$             \\ 
&  & Appendix A.4 &  Appendix A.4 &  Appendix A.4 &  Appendix A.4 \\
\cline{2-6} 
& \multirow{2}{*}{Machine Learner}  &     $\widehat\theta_V^{\text{eif$_1$-ml-s}}(a,a^*)$ & $\widehat\tau_V^{\text{eif$_1$-ml-s}}$          &        $\widehat\theta_V^{\text{eif$_2$-ml-s}}(a,a^*)$ & $\widehat\tau_V^{\text{eif$_2$-ml-s}}$             \\
&  &  Appendix A.4 &  Appendix A.4 &  Appendix A.4 &  Appendix A.4 \\
                               \hline
\end{tabular}}
\end{table}

\section{Simulation studies}\label{sec:simulation}

Web Appendix B presents simulation studies to evaluate the performance of the proposed methods under both correctly specified and misspecified models. We consider two data-generating settings. The first is a general setting designed to empirically verify the asymptotic properties of our proposed estimators. The second mimics the characteristics of the RPS study, aiming to assess the  performance of the methods in a context similar to our application. Across both settings, the proposed semiparametric doubly robust estimators and machine learning methods perform as predicted by theory, exhibiting minimal bias in scenarios where consistency is expected. When the number of clusters is small (as in RPS study), however, the stabilized machine learning estimators (including $\widehat{\zeta}^{\text{eif$_1$-ml-s}}$ and $\widehat{\zeta}^{\text{eif$_2$-ml-s}}$) may yield anti-conservative confidence intervals, whereas the remaining estimators appear less affected.

\section{Application to the RPS cluster-randomized trial}\label{sec:rps_app}


We assess the role of household dietary diversity in mediating the treatment effect from the cash-transfer program on child nutritional status. The mediator is measured by a household dietary diversity score (ranged between 0 and 12), and outcome is measured by child height-by-age $z$-score among children aged between 6--35 months (mean $-$1.7, standard deviation 1.2). Both the mediator and outcome were measured in a follow-up survey conducted two years after the intervention, and our mediation analysis assumes a biological mechanism whereby diet can influence height.  Due to data availability, we adjust for three individual-level baseline covariates: mother's educational level, mother's literacy, and highest education in household, which may not be sufficient to fully ensure Assumption \ref{assum:ignorability}; therefore, this analysis should be viewed primarily as an illustrative application. Among the 449 children, 16 had missing baseline covariate values, which were imputed using a single imputation procedure based on the \texttt{mice} package in R. 
Simulation suggests that the proposed methods perform satisfactorily with data similar to the RPS study, but stabilized machine learning estimators may yield anti-conservative confidence intervals due to the small sample size.  

\begin{sidewaysfigure}[ht]
\begin{center}
\includegraphics[width=0.9\textwidth]{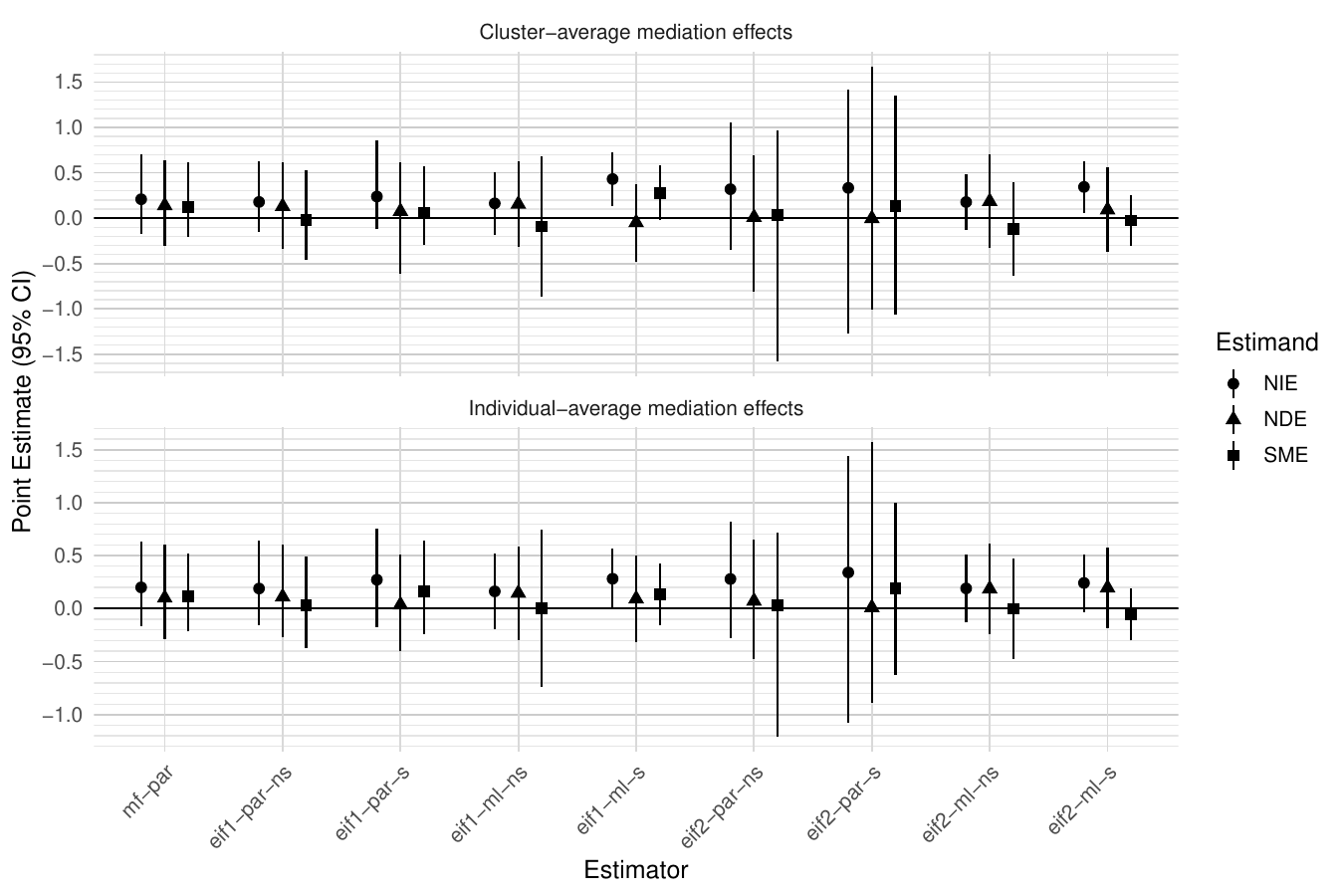}
\end{center}
\caption{Mediation of the effect of a cash-transfer program on child height-by-age $z$-score by household dietary divsersity score, the \textit{Red de Protecci\'on Social} pilot in Nicaragua, 2000--2002. Among the nine estimators, `mf-par' denotes the mediation functional estimator described in Section \ref{sec:nuisance}, while the remaining eight are EIF-induced estimators, with their definitions summarized  in Table \ref{tab:estimator_summary}.}
\label{fig:res_RPS}
\end{sidewaysfigure}

Figure \ref{fig:res_RPS} presents the NIE, NDE, and SME estimates on the difference scale. All methods yield positive  estimates for the NIE, although most of the corresponding 95\% confidence intervals include zero. Notably, the magnitude of the NIE is substantial relative to the NDE across all estimators. For instance, we observe $\widehat{\text{NIE}}_{C}^{\text{eif$1$-ml-ns}} = 0.18$ (95\% CI: [$-0.13, 0.49$]) and $\widehat{\text{NDE}}_{C}^{\text{eif$_1$-ml-ns}} = 0.18$ (95\% CI: [$-0.33, 0.70$]), suggesting that 50\% of TE is mediated through household dietary diversity. This provides suggestive evidence that the cash-transfer program may improve children’s nutrition partly by enhancing household dietary diversity. As a caveat, the two stabilized machine learning estimators ($\widehat{\text{NIE}}_C^{\text{eif$_1$-ml-s}}$ and $\widehat{\text{NIE}}_C^{\text{eif$_2$-ml-s}}$) indicate statistically significant NIEs, which may, however, due to anti-conservative confidence intervals arising from the small sample size. Furthermore, most SME estimates are close to zero, with wide 95\% confidence intervals including zero. This pattern suggests that child nutritional status is primarily influenced by each household's own dietary practices, with little impact from social interactions or dietary spillovers across households.

\section{Discussion}\label{sec:discussion}

{Assumptions \ref{assum:ignorability} and \ref{assum:no_icc} are not verifiable from the observed data. If Assumption \ref{assum:ignorability} fails to hold, our estimators may not be causally interpretable. An interpretable sensitivity analysis framework under violation of Assumption \ref{assum:ignorability} requires future work. Violation of Assumption \ref{assum:no_icc} also affects the causal interpretation. Interestingly, Remark 6 in Supplementary Material provides an alternative causal interpretation of SME and IME under the interventional mediation framework without relying on Assumption \ref{assum:no_icc}.  
It may also be of interest to study multiple mediators in CRTs. Recently, \cite{ohnishi2024bayesian} developed a Bayeisan nonparametric g-formula to assess the indirect and spillover mediation effects with unstructured multiple mediators in CRTs. However, semiparametric efficient estimators remain unavailable} 

Our nuisance function models assume within-cluster exchangeability, implemented through a Gaussian copula with an exchangeable correlation structure and a common regression model with symmetric summary statistics. Exploring alternative approaches that further simplify estimation of the joint mediator density $\kappa(a,\bmm, \bm c, n)$ is an important direction for future work. Moreover, although the exchangeability assumption is commonly invoked in analyzing CRTs, it may not hold when unobserved individual-level characteristics are associated with outcomes in ways that are not adequately captured by the current modeling framework. In such settings, the proposed estimators may be biased, and relaxing the exchangeability modeling assumption also merits future research.

\section*{Acknowledgement}
Research in this article was supported by the Patient-Centered Outcomes Research Institute\textsuperscript{\textregistered} (PCORI\textsuperscript{\textregistered} Award ME-2023C1-31350). All statements expressed in this article are solely those of the authors and do not necessarily reflect the views of PCORI\textsuperscript{\textregistered}.

\section*{Supplementary Materials}
Supplementary Material (available at \url{https://www.chaochengstat.com}) includes web appendices, an additional application study, and reproducible R code for analyses in Sections \ref{sec:simulation}--\ref{sec:rps_app}. The R package \texttt{DRmediateCRT} (available at \url{https://github.com/chaochengstat/DRmediateCRT}) is developed to implement the proposed approaches. 

\section*{Data Availability}

Data from the \textit{Red de Protección Social} (RPS) study are publicly available at \cite{DVN/EYQVWL_2005}.





\bibliographystyle{biom}
\bibliography{paper-ref}

\begin{thebibliography}{}

\bibitem[\protect\citeauthoryear{Bickel, Klaassen, Bickel, Ritov, Klaassen,
  Wellner, and Ritov}{Bickel et~al.}{1993}]{bickel1993efficient}
Bickel, P.~J., Klaassen, C.~A., Bickel, P.~J., Ritov, Y., Klaassen, J.,
  Wellner, J.~A., and Ritov, Y. (1993).
\newblock {\em Efficient and adaptive estimation for semiparametric models}.
\newblock Springer.

\bibitem[\protect\citeauthoryear{Charters, Kaufman, and Nandi}{Charters
  et~al.}{2023}]{charters2023causal}
Charters, T.~J., Kaufman, J.~S., and Nandi, A. (2023).
\newblock A causal mediation analysis for investigating the effect of a
  randomized cash-transfer program in nicaragua.
\newblock {\em American Journal of Epidemiology} {\bf 192,} 111--121.

\bibitem[\protect\citeauthoryear{Chernozhukov, Chetverikov, Demirer, Duflo,
  Hansen, Newey, and Robins}{Chernozhukov
  et~al.}{2018}]{chernozhukov2018double}
Chernozhukov, V., Chetverikov, D., Demirer, M., Duflo, E., Hansen, C., Newey,
  W., and Robins, J. (2018).
\newblock {Double/debiased machine learning for treatment and structural
  parameters}.
\newblock {\em The Econometrics Journal} {\bf 21,} C1--C68.

\bibitem[\protect\citeauthoryear{Cruces, Tortarolo, and Vazquez-Bare}{Cruces
  et~al.}{2025}]{cruces2025design}
Cruces, G., Tortarolo, D., and Vazquez-Bare, G. (2025).
\newblock Design of partial population experiments with an application to
  spillovers in tax compliance.
\newblock {\em Review of Economics and Statistics} pages 1--45.

\bibitem[\protect\citeauthoryear{Davezies, d'Haultfoeuille, and
  Foug{\`e}re}{Davezies et~al.}{2009}]{davezies2009identification}
Davezies, L., d'Haultfoeuille, X., and Foug{\`e}re, D. (2009).
\newblock Identification of peer effects using group size variation.
\newblock {\em The Econometrics Journal} {\bf 12,} 397--413.

\bibitem[\protect\citeauthoryear{D{\'\i}az, Hejazi, Rudolph, and van
  Der~Laan}{D{\'\i}az et~al.}{2021}]{diaz2021nonparametric}
D{\'\i}az, I., Hejazi, N.~S., Rudolph, K.~E., and van Der~Laan, M.~J. (2021).
\newblock Nonparametric efficient causal mediation with intermediate
  confounders.
\newblock {\em Biometrika} {\bf 108,} 627--641.

\bibitem[\protect\citeauthoryear{Field and Welsh}{Field and
  Welsh}{2007}]{field2007bootstrapping}
Field, C.~A. and Welsh, A.~H. (2007).
\newblock Bootstrapping clustered data.
\newblock {\em Journal of the Royal Statistical Society Series B: Statistical
  Methodology} {\bf 69,} 369--390.

\bibitem[\protect\citeauthoryear{Hudgens and Halloran}{Hudgens and
  Halloran}{2008}]{hudgens2008toward}
Hudgens, M.~G. and Halloran, M.~E. (2008).
\newblock Toward causal inference with interference.
\newblock {\em Journal of the American Statistical Association} {\bf 103,}
  832--842.

\bibitem[\protect\citeauthoryear{{International Food Policy Research
  Institute}}{{International Food Policy Research
  Institute}}{2005}]{DVN/EYQVWL_2005}
{International Food Policy Research Institute} (2005).
\newblock {Nicaragua, Red de Protección Social (RPS) Evaluation Dataset,
  2000-2002, Harvard Dataverse, Version 4, DOI: 10.7910/DVN/EYQVWL}.

\bibitem[\protect\citeauthoryear{Jiang, Imai, and Malani}{Jiang
  et~al.}{2023}]{jiang2023statistical}
Jiang, Z., Imai, K., and Malani, A. (2023).
\newblock Statistical inference and power analysis for direct and spillover
  effects in two-stage randomized experiments.
\newblock {\em Biometrics} {\bf 79,} 2370--2381.

\bibitem[\protect\citeauthoryear{Kahan, Li, Copas, and Harhay}{Kahan
  et~al.}{2023}]{kahan2023estimands}
Kahan, B.~C., Li, F., Copas, A.~J., and Harhay, M.~O. (2023).
\newblock Estimands in cluster-randomized trials: choosing analyses that answer
  the right question.
\newblock {\em International Journal of Epidemiology} {\bf 52,} 107--118.

\bibitem[\protect\citeauthoryear{Masarotto and Varin}{Masarotto and
  Varin}{2012}]{masarotto2012gaussian}
Masarotto, G. and Varin, C. (2012).
\newblock Gaussian copula marginal regression.
\newblock {\em Electronic Journal of Statistics} {\bf 6,} 1517--1549.

\bibitem[\protect\citeauthoryear{Ogburn, Sofrygin, Diaz, and Van~der
  Laan}{Ogburn et~al.}{2022}]{ogburn2022causal}
Ogburn, E.~L., Sofrygin, O., Diaz, I., and Van~der Laan, M.~J. (2022).
\newblock Causal inference for social network data.
\newblock {\em Journal of the American Statistical Association} pages 1--15.

\bibitem[\protect\citeauthoryear{Ohnishi and Li}{Ohnishi and
  Li}{2025}]{ohnishi2024bayesian}
Ohnishi, Y. and Li, F. (2025).
\newblock A bayesian nonparametric approach to mediation and spillover effects
  with multiple mediators in cluster-randomized trials.
\newblock {\em Journal of the American Statistical Association} pages 1--20.

\bibitem[\protect\citeauthoryear{Park and Kaplan}{Park and
  Kaplan}{2015}]{park2015bayesian}
Park, S. and Kaplan, D. (2015).
\newblock Bayesian causal mediation analysis for group randomized designs with
  homogeneous and heterogeneous effects: Simulation and case study.
\newblock {\em Multivariate Behavioral Research} {\bf 50,} 316--333.

\bibitem[\protect\citeauthoryear{Phillips, Van Der~Laan, Lee, and
  Gruber}{Phillips et~al.}{2023}]{phillips2023practical}
Phillips, R.~V., Van Der~Laan, M.~J., Lee, H., and Gruber, S. (2023).
\newblock Practical considerations for specifying a super learner.
\newblock {\em International Journal of Epidemiology} {\bf 52,} 1276--1285.

\bibitem[\protect\citeauthoryear{Tchetgen~Tchetgen and
  Shpitser}{Tchetgen~Tchetgen and Shpitser}{2012}]{tchetgen2012semiparametric}
Tchetgen~Tchetgen, E.~J. and Shpitser, I. (2012).
\newblock Semiparametric theory for causal mediation analysis: efficiency
  bounds, multiple robustness, and sensitivity analysis.
\newblock {\em Annals of Statistics} {\bf 40,} 1816.

\bibitem[\protect\citeauthoryear{Van~der Laan, Rose, et~al\mbox{.}}{Van~der
  Laan et~al.}{2011}]{van2011targeted}
Van~der Laan, M.~J., Rose, S., et~al. (2011).
\newblock {\em Targeted learning: causal inference for observational and
  experimental data}, volume~4.
\newblock Springer.

\bibitem[\protect\citeauthoryear{VanderWeele}{VanderWeele}{2010}]{vanderweele2010direct}
VanderWeele, T.~J. (2010).
\newblock Direct and indirect effects for neighborhood-based clustered and
  longitudinal data.
\newblock {\em Sociological Methods \& Research} {\bf 38,} 515--544.

\bibitem[\protect\citeauthoryear{VanderWeele, Hong, Jones, and
  Brown}{VanderWeele et~al.}{2013}]{vanderweele2013mediation}
VanderWeele, T.~J., Hong, G., Jones, S.~M., and Brown, J.~L. (2013).
\newblock Mediation and spillover effects in group-randomized trials: a case
  study of the 4rs educational intervention.
\newblock {\em Journal of the American Statistical Association} {\bf 108,}
  469--482.

\bibitem[\protect\citeauthoryear{Wang, Park, Small, and Li}{Wang
  et~al.}{2023}]{wang2023model}
Wang, B., Park, C., Small, D.~S., and Li, F. (2023).
\newblock Model-robust and efficient covariate adjustment for
  cluster-randomized experiments.
\newblock {\em Journal of the American Statistical Association} pages 1--28.

\bibitem[\protect\citeauthoryear{Zhou}{Zhou}{2022}]{zhou2022semiparametric}
Zhou, X. (2022).
\newblock Semiparametric estimation for causal mediation analysis with multiple
  causally ordered mediators.
\newblock {\em Journal of the Royal Statistical Society Series B: Statistical
  Methodology} {\bf 84,} 794--821.

\end{thebibliography}






\label{lastpage}

\end{document}